\documentclass[namedreferences]{kluwer}    % Specifies the document style.
\usepackage{graphicx}

\begin{document}                                                                                   
\begin{article}
\begin{opening}         
\title{Kinematics of diffuse elliptical galaxies}
\author{Ph. \surname{Prugniel} \email{prugniel@obs.univ-lyon1.fr}}  
\author{F.  \surname{Simien}}  
\runningauthor{Prugniel \& Simien}
\runningtitle{Diffuse ellipticals}
\institute{Observatoire de Lyon (CRAL)}

\date{September 30, 2002}

\begin{abstract}
New observations show that most of the flattened diffuse elliptical (dE) 
galaxies are
essentially isotropic rotators\@. This supports the idea that dEs have evolved
from fast rotating dIrr systems or late-type spirals after their gas was
expelled by SN-driven winds and stripped by ram pressure against the 
intergalactic medium\@.
\end{abstract}

\end{opening}            

%------------------------------------------------------------------------------
\section{Observational material}  
We have published kinematical observations of 20 dEs in \citeauthor{Simien02}
\shortcite{Simien02}\@. They were obtained at the CARELEC long-slit 
spectrograph attached to the 2-m telescope of the Observatoire de 
Haute-Provence\@. Figure 1 presents two typical velocity 
profiles: those of NGC 205 (extending out to 5 arcmin), and IC 3328, a 
Virgo-Cluster dE\@. 

The kinematical profiles, extending beyond one effective radius, reveal 
significant rotation in all our candidate objects\@. A similar 
result is presented in \citeauthor{Pedraz02} \shortcite{Pedraz02}\@. 

From these and other sources, we have selected 14 galaxies with reliable 
kinematical measurements\@. We did not include four of the faint Virgo dwarfs 
observed by \citeauthor{Geha02} \shortcite{Geha02}, where no rotation was 
detected, because the profiles extend to only a fraction of the effective 
radius; experience from past observations reveals that extrapolations of 
central profiles is hazardous\@. 

%------------------------------------------------------------------------------
\section{Discussion of the kinematical data}
For more than 10 years, dE galaxies were though to be flattened by
anisotropy \cite{Bender90}\@. Because new observations extend farther out,
more rotation can be detected\@.

Since dEs feature roughly exponential photometric profiles, we cannot
apply directly the $V/\sigma$ anisotropy test \cite{Binney78}, which is
adapted to the case of $r^{1/4}$ profiles\@. We thus developed a simple 
isotropic dynamical model as in \citeauthor{Loyer98} \shortcite{Loyer98}, 
but for a $r^{1/n}$ profile \cite{Sersic68}\@. Letting $a$ be the semi-major 
axis of an isophote 
(with $a_{\rm e}$ corresponding to the effective isophote), Figure 2 presents 
typical normalized rotation curves for $n=1$ and $n=4$\@. The model shows that 
the maximum rotation is expected at $a\sim 2\,a_{\rm e}$ for a dE ($n=1$ or 2), 
compared to $a\sim 0.5\, a_{\rm e}$ for an elliptical ($n=4$)\@. And for a dE, 
the expected $V(a_{\rm e})/\sigma_0$ ratio (where $\sigma_0$ is the peak 
central velocity dispersion) is 10$\sim$20 \% larger than for a normal E\@.

%..............................................................................
\begin{figure}
\centerline{\includegraphics[width=25pc]{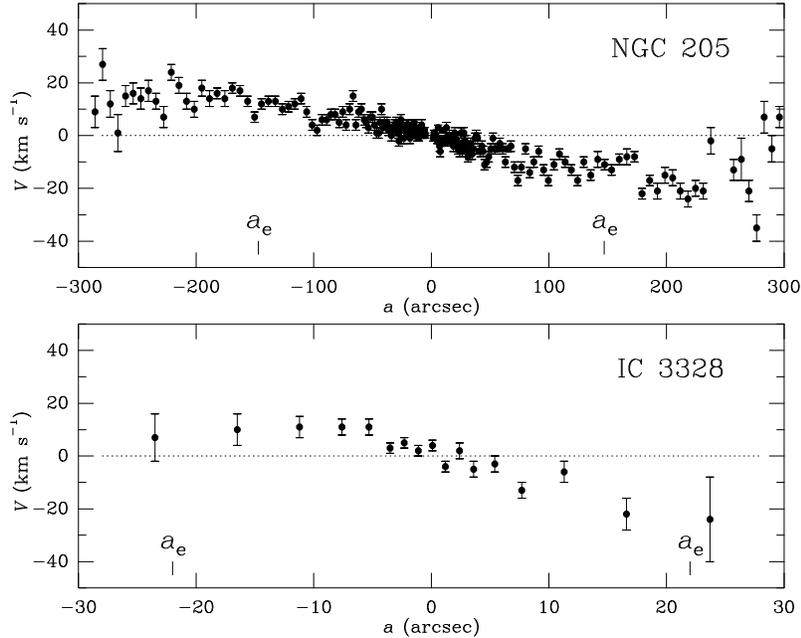}}
\caption{Typical stellar rotation curves, along the major axis, for two 
dE galaxies (from Simien \& Prugniel, 2002).}
\end{figure}
%..............................................................................

Figure 3 shows the kinematic test for our data supplemented by other sources\@.
Most galaxies
are compatible with being rotationally supported\@. The two flattest objects,
IC 3393 and IC 3773, bear large error bars but do not seem incompatible
with isotropy\@. Other interesting cases are NGC 205 and IC 794\@. Despite the
significant rotation we have found, the former is unlikely to be isotropic;
and the latter is certainly anisotropic\@.

%------------------------------------------------------------------------------
\section{The origin of dE galaxies}  
Several similarities do exist between E and dE galaxies: a) there is a 
continuum in the surface brightness {\it versus} luminosity diagram 
\cite{Prugniel94}, b) there is a transition in the shape of the 
photometric profile and, c) on taking into account the difference in stellar
populations \cite{Prugniel96}, the dE galaxies lie on the Fundamental Plane
of ellipticals\@. These characteristics, however, reflect equilibrium relations
and do not necessarily imply a common origin\@. Conversely, the difference in 
clustering properties of the two classes is a clue for different origins\@.

The new kinematical observations imply that the dE progenitors are probably
rotating dIrr systems and low-mass, late-type spirals\@. As in the Dekel et al. 
\shortcite{Dekel86} scenario, the intrinsic low-mass of the progenitors may 
result
in an important mass loss due to the SN-driven winds, and this may explains 
their low
metallicity\@. In addition, the ram pressure against the hot intergalactic 
medium strips off the residual neutral gas\@. Dynamical harassment from 
encounters with massive galaxies, which predicts anisotropic objects 
\cite{Moore98}, does not appear to represent a key factor for these dEs, but
ram pressure stripping of the gas may have helped to stop the star formation\@.

%..............................................................................
\begin{figure}
\centerline{\includegraphics[width=18pc]{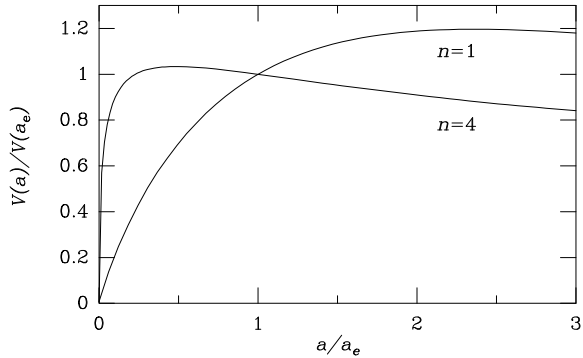}}
\caption{Model rotation curves for an oblate, isotropic rotator with 
$r^{1/n}$ photometric profile and intrinsic ellipticity $\epsilon=0.3$.}
\end{figure}
%..............................................................................

A few interesting objects deserve additional comments\@.
NGC 205 may own its anisotropy to its
interaction with M 31 \cite{Mayer01}\@. But we note that the relevance of the 
diagnostic may be questioned, since the observed kinematics fail to fit the
isotropic model with a constant {\it M/L} ratio\@.
IC 794 is particularly interesting\@. As noted by Pedraz et al. 
\shortcite{Pedraz02},
it is strongly anisotropic, young, and metal rich\@. These characteristics make
it a good candidate for being a tidal dwarf issued from enriched material
after a major collision involving a spiral\@. We have no information on the
stellar content of IC 3344, the other anisotropic object of 
\citeauthor{Geha02}, which may be similar to IC 794 (both lie in center of 
the Virgo cluster)\@.

The other galaxies where \citeauthor{Geha02} did not detect rotation are 
fainter; they may have been more sensitive to environment and predisposed
to anisotropy\@. But observations extending to larger radii are needed to 
conclude\@.

%..............................................................................
\begin{figure}
\centerline{\includegraphics[width=28pc]{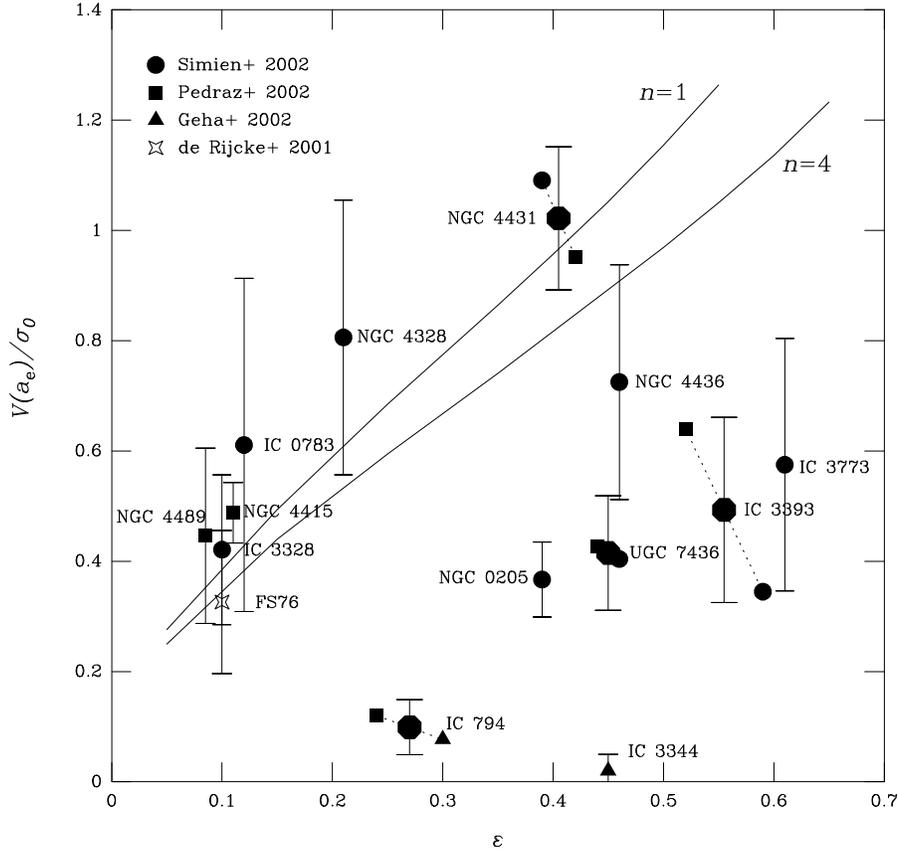}}
\caption{The $V/\sigma$ kinematic test for a sample of dE galaxies.}
\end{figure}
%..............................................................................

%------------------------------------------------------------------------------

\end{article}

\begin{thebibliography}{}

\bibitem[\protect\citeauthoryear{Bender \& Nieto}{1990}]{Bender90} Bender, R.,
Nieto, J.-L.: 1990, A\&A, 239, 97-112 

\bibitem[\protect\citeauthoryear{Binney}{1978}]{Binney78}
Binney, J.: 1978, MNRAS, 183, 501-514

\bibitem[\protect\citeauthoryear{Dekel \& Silk}{1986}]{Dekel86}
Dekel, A., Silk, J.: 1986, ApJ, 303, 39-55

\bibitem[\protect\citeauthoryear{de Rijcke et al.}{2001}]{deRijcke01} de 
Rijcke, S., Dejonghe, H., Zeilinger, W.~W., Hau, G.~K.~T.: 2001, ApJ, 559, 
L21-L24 

\bibitem[\protect\citeauthoryear{Geha et al.}{2002}]{Geha02} Geha, M.,
Guhathakurta, P., van der Marel, R.: 2002, preprint, astro/ph 0206153 

\bibitem[\protect\citeauthoryear{Loyer et al.}{1998}] {Loyer98} Loyer, E., 
Simien, F., Michard, R., Prugniel, Ph.: 1998, A\&A, 334, 805-813

\bibitem[\protect\citeauthoryear{Mayer et al.}{2001}]{Mayer01} Mayer, L., 
Governato, F., Colpi, M., Moore, B., Quinn, T., Wadsley, J., Stadel, J., 
Lake, G.: 2001, ApJ, 559, 754-784

\bibitem[\protect\citeauthoryear{Moore et al.}{1998}]{Moore98} 
Moore, B., Lake, G., Katz, N.: 1998, ApJ, 495, 139-151

\bibitem[\protect\citeauthoryear{Pedraz et al.}{2002}]{Pedraz02} Pedraz, S., 
Gorgas, J., Cardiel, N., S{\' a}nchez-Bl{\' a}zquez, P., Guzm{\' a}n, R.: 
2002, MNRAS, 332, L59-L63

\bibitem[\protect\citeauthoryear{Prugniel}{1994}] {Prugniel94} 
Prugniel, Ph.: 1994,in G. Meylan and Ph. Prugniel, editors, {\it Proceedings 
of an ESO/OHP Workshop on Dwarf galaxies}\@. ESO, Garching, p. 171

\bibitem[\protect\citeauthoryear{Prugniel \& Simien}{1996}] {Prugniel96} 
Prugniel, Ph., Simien, F.: 1996, A\&A, 309, 749-759

\bibitem[\protect\citeauthoryear{S\'ersic}{1968}]{Sersic68} S\'ersic, J.~L.: 
1968, Atlas de Galaxias Australes, Observatorio de C\'ordoba

\bibitem[\protect\citeauthoryear{Simien \& Prugniel}{2002}]{Simien02} 
Simien, F., Prugniel, P.: 2002, A\&A, 384, 371-382 

\end{thebibliography}
\end{document}